\documentstyle[prl,aps,tighten,preprint,epsf]{revtex}

\begin{document}

\bibliographystyle{prsty}

\widetext

\draft
\title{A novel intermediate phase in $S=2$ antiferromagnetic
chains with uniaxial anisotropy}
\author{Masaki Oshikawa\cite{oshiemail}}
\address{
Department of Applied Physics, University of Tokyo,
Bunkyo-ku, Tokyo 113, Japan}
\author{Masanori Yamanaka\cite{yamanakaemail}}
\address{
Institute for Solid State Physics, University of Tokyo
Minato-ku, Tokyo 106, Japan}
\author{Seiji Miyashita}
\address{Department of Earth and Space Science,
Osaka University, Machikaneyamacho, Toyonaka 560, Japan}

\date{\today}
\maketitle

\begin{abstract}
We study ground state properties of the $S=2$ quantum
antiferromagnetic chain with a uniaxially anisotropic Hamiltonian:
$ H = \sum_{j} [ \mbox{\boldmath $S$}_{j} \cdot \mbox{\boldmath $S$}_{j+1}
                 + D (S^{z}_{j})^2 ]
$
by a Monte Carlo calculation.
While it has been reported that a string order parameter
vanishes exponentially at the isotropic Heisenberg point ($D=0$),
we found it tends to remain finite in the thermodynamic limit
around $D=1.2$.
This implies that the model for $S=2$ has a novel
intermediate phase between the Haldane phase
and the large-$D$ phase,
as was anticipated from the earlier argument based on the
Valence-Bond-Solid picture.
\end{abstract}

\narrowtext

Since Haldane predicted~\cite{Haldane:conj1,Haldane:conj2}
a qualitative difference
between quantum antiferromagnetic chains
with integer and half-odd-integer spins,
much work has been devoted to examine his prediction.
Now it is widely believed based on
many theoretical, numerical and experimental studies.
(For a review, see~\cite{Affleck:reviewHaldane}).
In the present letter, we report numerical results which indicate that
integer spin chains with $S>1$ exhibit much richer structures
than expected, and are still full of fascinating surprises.

Exactly solvable models of integer spin chains,
which exhibit the properties predicted by Haldane,
were introduced by
Affleck, Kennedy, Lieb and Tasaki~\cite{AKLT}.
It was found that
the exact ground state --- Valence-Bond-Solid (VBS) state ---
of the above model with $S=1$
possesses two peculiar properties:
the presence of edge states in a finite open chain,
and the existence of a hidden antiferromagnetic order.
The hidden antiferromagnetic order can be measured by the
den Nijs-Rommelse string order parameter~\cite{dNR89}
\begin{equation}
\label{eq:dNRorder}
  {\cal O}^{\alpha}_{\rm str} = \lim_{|i-j| \rightarrow \infty}
    \langle S^{\alpha}_{i} \exp{(i \pi \sum_{i \leq k < j} S^{\alpha}_{k})}
            S^{\alpha}_{j} \rangle .
\end{equation}
Later it was
argued~\cite{dNR89,Hal:sto,GirvinArovas,HatsugaiKohmoto:str,Kennedy:edge,Miya:edge1,Miya:edge2}
that the above two properties are characteristic to general $S=1$ chains
which are in the Haldane phase including the standard $S=1$
Heisenberg model.
Kennedy and Tasaki~\cite{KennedyTasaki:hid} showed that
these properties can be viewed as
consequences of spontaneous breakdown of a
hidden $Z_2 \times Z_2$ symmetry.

One of us (M.O.)~\cite{MOVBS} extended the notion of
hidden $Z_2 \times Z_2$
symmetry to chains with arbitrary integer spin.
Earlier,
Affleck and Haldane~\cite{AffHal87} made a field theory argument
to predict that the quantum antiferromagnetic chain with
a bond alternation undergoes $2S$ successive phase transitions
when the magnitude of the alternation is changed.
Based on VBS-type models, he argued in~\cite{MOVBS} that
breaking of the hidden $Z_2 \times Z_2$ symmetry is important in
these successive dimerization transitions.
Moreover, he conjectured unexpected successive transitions in
integer $S>1$ quantum antiferromagnetic chains with uniaxial
anisotropy.

We summarize the conjecture for $S=2$ below.
Let us consider a quantum spin chain with the Hamiltonian
\begin{equation}
  \label{eq:Dham}
  H = \sum_{j} [ J \mbox{\boldmath $S$}_{j} \cdot \mbox{\boldmath $S$}_{j+1}
                 + D (S^{z}_{j})^2 ] .
\end{equation}
We take $J$ as the unit of energy and put it unity hereafter.
For $S=1$, the phase structure has been thoroughly
investigated~\cite{dNR89,KennedyTasaki:hid,BotetJulienKolb,SchulzZiman86,Hal:sto}.
A finite region in $D$ including the Heisenberg point ($D=0$)
belongs to the Haldane gap phase which is characterized by a
spontaneous breakdown of the hidden $Z_2 \times Z_2$ symmetry.
When $D$ is sufficiently large, it belongs to a different
phase called the large-$D$ phase,
in which the hidden symmetry is unbroken.

\begin{figure}[htbp]
  \begin{center}
    \leavevmode
    \epsfbox{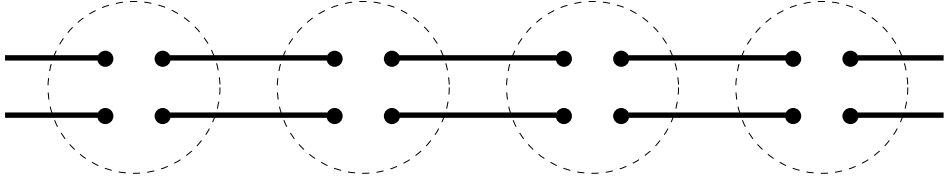}
  \end{center}
\caption{The standard VBS state for $S=2$.
A small solid circle denotes
a spin-$1/2$, and a solid line denotes a valence bond (singlet of two
spin-$1/2$). A dotted circle represents the symmetrization of
four spin-$1/2$'s at each site to form a spin-2.}
\label{fig:s2vbs}
\end{figure}

\begin{figure}[htbp]
  \begin{center}
    \leavevmode
    \epsfbox{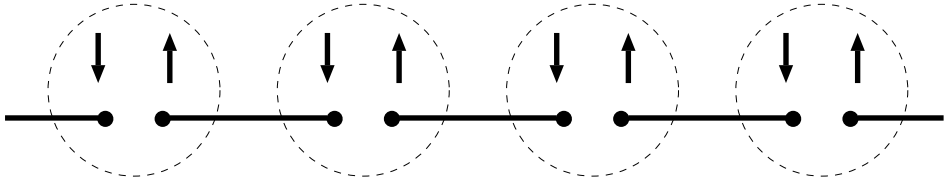}
  \end{center}
\caption{The intermediate-$D$ VBS state for $S=2$.
An up (down) arrow denotes an
 up (down) spin-$1/2$. (cf. Fig.~\protect\ref{fig:s2vbs}) }
\label{fig:idvbs}
\end{figure}

\begin{figure}[htbp]
  \begin{center}
    \leavevmode
    \epsfbox{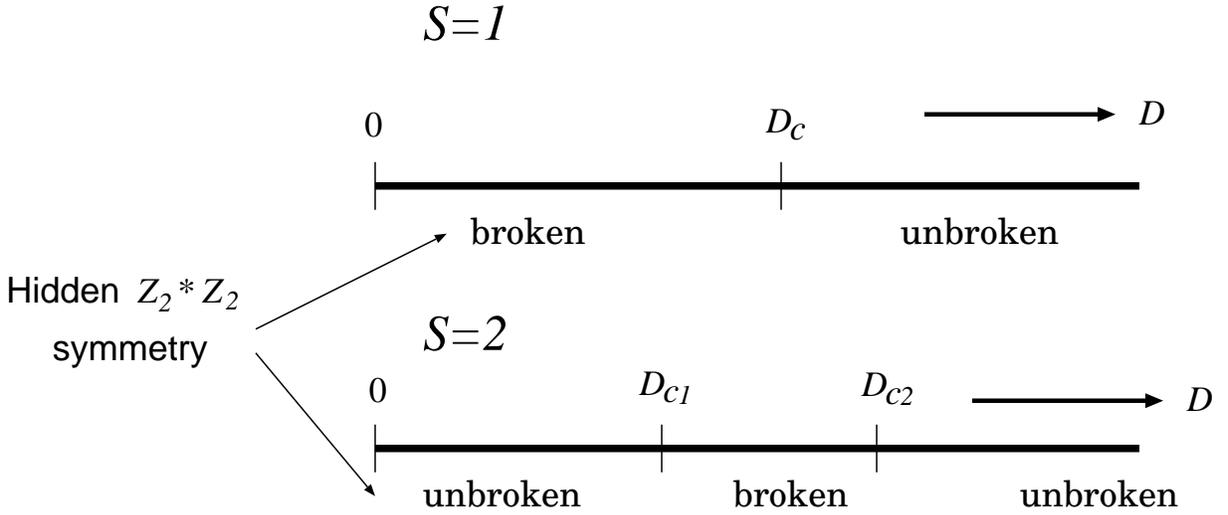}
  \end{center}
\caption{The established phase structure for $S=1$, and
the conjectured phase structure for $S=2$
for the Hamiltonian~(2).}
\label{fig:s12phased}
\end{figure}

For $S=2$, the ground state at the isotropic Heisenberg point ($D=0$)
is expected to be similar to the $S=2$ VBS state (Fig.~\ref{fig:s2vbs}).
In this state,
the den Nijs-Rommelse string order parameter~(\ref{eq:dNRorder})
is exactly evaluated~\cite{MOVBS} to be zero; the hidden
$Z_2 \times Z_2$ symmetry is unbroken here.
When $D \rightarrow \infty$, the ground state tends toward
the simple tensor product state with $S^z =0$ for each site.
The hidden symmetry is also unbroken in this state and
this is similar to the large-$D$ phase for $S=1$.
However, in an intermediate range of $D$, the ground state
may become similar to the incomplete
VBS-type state as in Fig.~\ref{fig:idvbs}.
Here the hidden symmetry is shown~\cite{MOVBS} to be broken.
 From the symmetry argument we expect that
there exist a new ``intermediate-$D$ phase'' between the
Haldane phase and the large-$D$ phase
for $S=2$, in contrast to the well-established two phases for $S=1$
(Fig.~\ref{fig:s12phased}).

On the other hand, the validity of the VBS-model description is not
clear for $S>1$.
While the VBS-model provides a simple way of understanding
the difference between integer and half-odd-integer spins,
the Hamiltonian of the model contains many extra terms
when $S>1$.
Actually a bosonization argument~\cite{Schulz86} concludes
that the qualitative feature of the phase diagram of an
anisotropic Heisenberg-type model is
universal for any integer spin, against the above conjecture.

In this letter, we numerically examine the existence of the
conjectured intermediate phase in the $S=2$ chain with
the Hamiltonian~(\ref{eq:Dham}).
Before presenting our results,
we briefly summarize previous studies on the $S=2$ spin chain.
Numerical confirmation of the Haldane's conjecture for the $S=2$
Heisenberg model ($D=0$) is done by several
authors~\cite{HatanoSuzuki93:S=2,Deiszetal93:spin,Meshkov93:spin,Nishiyamaetal95:S=2}.
All the results support the Haldane's conjecture:
the energy gap above the ground state is finite and
the spin correlation function decays with a finite correlation
length.
Furthermore, Meshkov~\cite{Meshkov93:spin} and
Nishiyama et al.~\cite{Nishiyamaetal95:S=2}
examined the den Nijs-Rommelse string correlation function
at the Heisenberg point.
They found that the string order seems exponentially vanishing;
the hidden $Z_2 \times Z_2$ symmetry is unbroken there.
Instead, Nishiyama et al. found a modified
string order parameter~\cite{MOVBS} remains finite.
(See also~\cite{Hatsugai:S=2str}).
These results qualitatively coincides with the properties of
the $S=2$ VBS state.

Let us also note
that it is trivial that the hidden $Z_2 \times Z_2$ symmetry
is unbroken at $D = \infty$.
In addition, applying the rigorous perturbation theory
by Kennedy and Tasaki~\cite{KennedyTasaki:hid},
it can be stated rigorously that the symmetry is unbroken for
sufficiently large but finite $D$.

Thus we concentrate to find a hidden $Z_2 \times Z_2$ symmetry
breaking in some finite value of $D$.
We performed a world-line
Quantum Monte Carlo calculation~\cite{Suzuki76:Trotter} using the
Lie-Trotter-Suzuki product formula
with the checker-board decomposition~\cite{Hirshetal:checkerboard}.
That is, we made an approximation to the partition function $Z$
for temperature $T$ as
\begin{equation}
\label{eq:decomposed}
  Z_n = \mbox{Tr}[( e^{-{H_A}/{(nT)}} e^{-{H_B}/{(nT)}})^n].
\end{equation}
Here we choose
$H_A = \sum_{j = {\rm odd}}  V_j$ ,
$H_B = \sum_{j = {\rm even}} V_j$ and
$V_j =   \mbox{\boldmath S}_{j} \cdot \mbox{\boldmath S}_{j+1}
                 + \frac{D}{2} ({S^{z}_{j}}^2 + {S^{z}_{j+1}}^2)$.
The approximate partition function
$Z_n$ approaches to the true partition function $Z$ as
$n \rightarrow \infty$.
We made calculations for several values
of finite $n$, and
then extrapolate the results to $n \rightarrow \infty$.
Inserting the sum of complete set of bases,
the decomposed formula~(\ref{eq:decomposed}) can be interpreted
as a classical spin system with a four-body interaction.
Monte Carlo flips for each plaquette
were performed with a heat-bath algorithm.
Although we also prepared global flips along the chain direction,
the acceptance ratio becomes very small when the system size
is large (over $100$).
For the study of the ground state,
we did not use global flips along the Trotter direction and
restricted the calculation into the $\sum S^z = 0$ subspace.

Scanning several values of $D$ by preliminary calculations,
we have chosen $D=1.2$ as a candidate for a point in the
intermediate phase.
We calculated spin chains with a periodic boundary
condition up to the system size $L=160$.
We investigate the quantities at low enough temperature,
below which the correlation function is temperature independent.
For actual calculation
we take $T=0.04$ for $L \leq 140$ and $T=0.02$ for $L=160$.
The Trotter numbers $n$ used for the calculation were
$48,64,72,80$ and $96$ for $L \leq 140$ and $96,128,144$ and
$192$ for $L=160$.

Typical data are obtained as follows.
To reduce the effect of the autocorrelation,
the measurements were done in every 10 Monte Carlo Steps (MCS).
In each configuration, we chose 800 points as positions of $i$
from the classical spin
lattice of size $L \times 2n$.
For each point,
we scan $j$ to the chain direction to measure
the den Nijs-Rommelse string correlation function (in $z$-axis)
\[
    \langle S^{z}_{i} \exp{(i \pi \sum_{i \leq k < j} S^{z}_{k})}
            S^{z}_{j} \rangle,
\]
and the $S^z$ correlation function
$ \langle S^{z}_{i} S^{z}_{j} \rangle $
as functions of $j-i$.
We performed the measurement during $1 \sim 5 \times 10^{6}$ MCS,
after $1 \times 10^{6} \sim 9 \times 10^{6}$ MCS of thermalization.

Error bars are calculated from the standard deviation of
averages during each $1 \times 10^{4}$ MCS.
The dependence on the Trotter number is extrapolated by the
least square method with the formula
$  A(n) = A_{0} + A_1/n^2 $,
where $A(n)$ is a physical quantity for the Trotter number $n$
and $A_0$ and $A_1$ are the unknown constants (fitting parameters).
We have checked that the extrapolation is stable
under different choices of Trotter numbers or
inclusion of the ${A_2}/{n^4}$ term.

\begin{figure}[htbp]
  \begin{center}
    \leavevmode
   \noindent
    \epsfbox{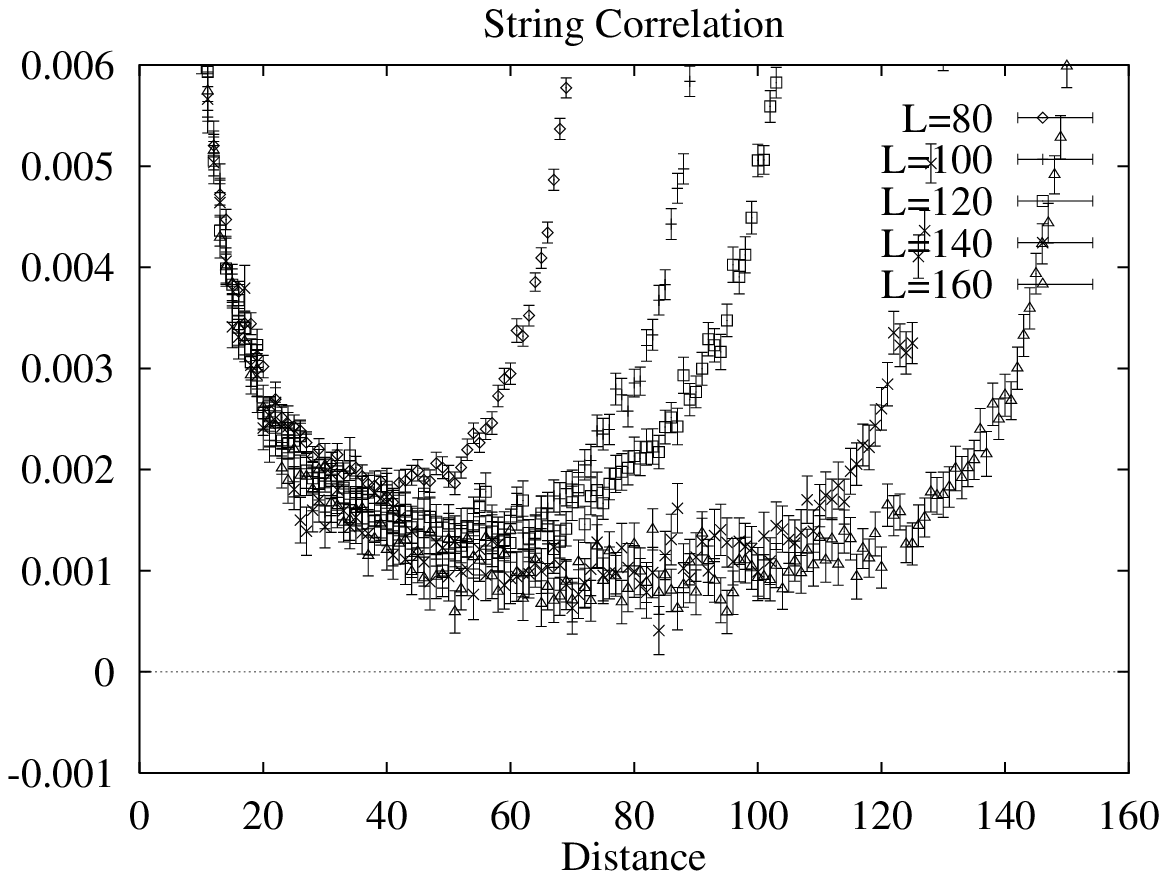}
\\
\bigskip
\bigskip
    \epsfbox{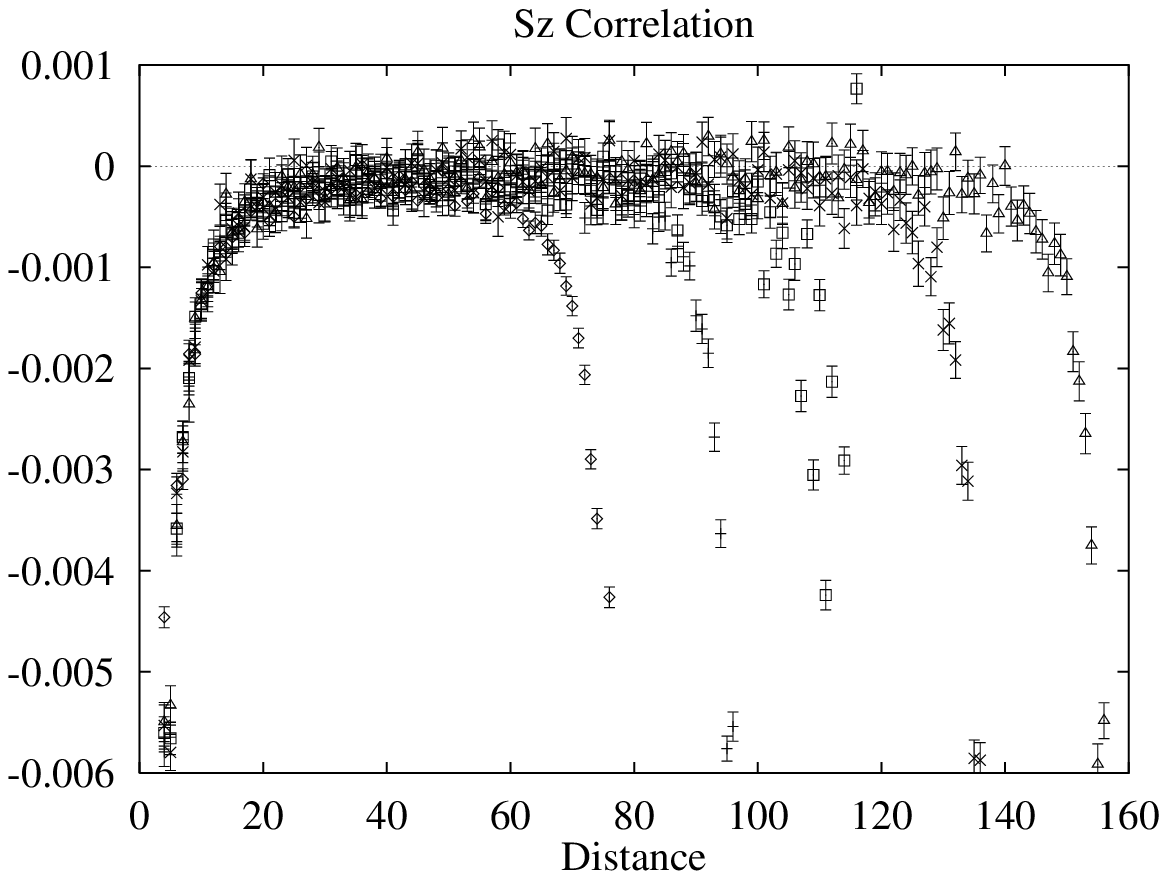}
  \end{center}
\caption{Extrapolated correlation functions for system sizes
$L= 80,100,120$ and $140$ at $D=1.2$.
The string correlation seems to show a plateau at $0.001$,
while the $S^z$ correlation is vanishing.}
\label{fig:D12corr}
\end{figure}

In Fig.~\ref{fig:D12corr}
we show the extrapolated correlation functions for
system sizes $L= 80, 100, 120, 140$ and $160$.
The expected antiferromagnetic part in $S^z$ correlation
is dominant only for nearest few sites and decays very rapidly.
Actually, in the figure
the major part of the $S^z$ correlation function is
not antiferromagnetic.
It is rather a negative correlation
which can be explained by the sum rule $\sum_j S^z_j =0$ in
the ground state.
Anyway, the $S^z$-correlation (or N\'{e}el correlation) decays to zero.
On the other hand,
the den Nijs-Rommelse string correlation
seems to show a plateau at a finite positive value around $0.001$.
Because the value is so small, we checked the thermal equilibrium
carefully by comparing the results for various length of the thermalization.
We also performed a lot of measurement in order to make the statistical
error less than $0.0001$.
We note that the string correlation is always positive,
which is contrasted to the oscillating behavior at the Heisenberg
point~\cite{Hatsugai:S=2str}.

It seems that the string correlation converges to a finite positive
value around $0.001$ in the thermodynamic limit,
in contrast to the vanishing N\'{e}el order.
However, there is a subtlety about the presence of the true long-range
string order.
The string correlation function is fit well with a power-law function
\begin{equation}
\label{eq:powerlaw}
0.12 x^{-1.3}
\end{equation}
within a range of the distance.
It is difficult to distinguish a true long-range order from
a power-law decay.

\begin{figure}[htbp]
  \begin{center}
    \noindent
    \leavevmode
    \epsfbox{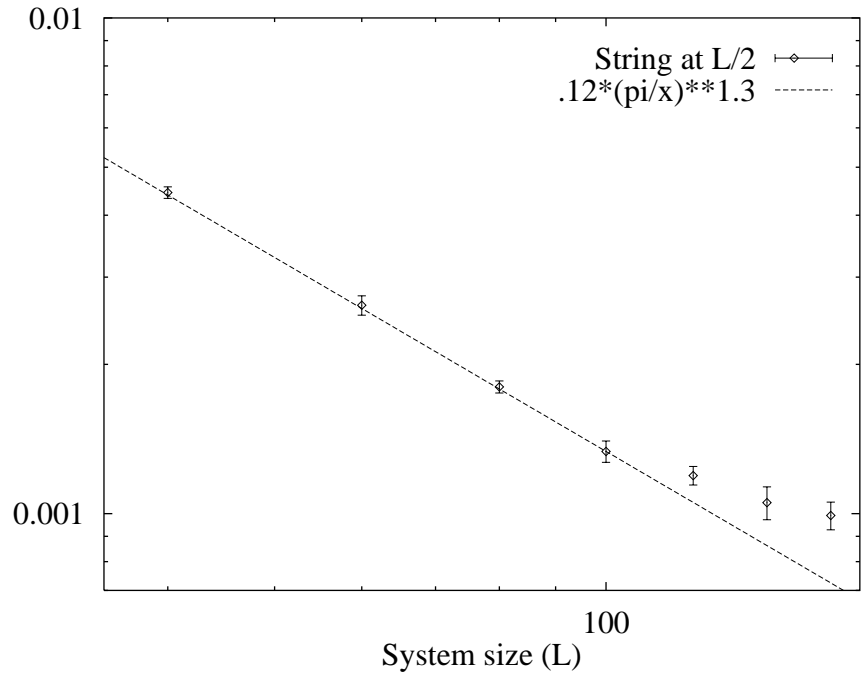}
  \end{center}
\caption{The value of the string correlation function at the middle point
as a function of the system size.
The straight line is a theoretical prediction assuming the conformal
invariance.}
\label{fig:fits}
\end{figure}

To examine this problem,
let us compare the string correlation data with the assumption that
the system is in a criticality with the conformal invariance.
(See for example~\cite{ItzyksonDrouffe}
for a review.
On applications to the finite size scaling of correlations,
see~\cite{Cardy:1986,KomaMizukoshi})
Under the assumption, string correlation corresponds to
the two-point correlation function of a spinless primary field $\phi$
with the conformal weight $h = \bar{h}$, whose
correlation function in an infinite system is given by
$  \langle \phi(x) \phi(y) \rangle = {C}/{|x - y|^{2h}} $
where $C$ is a constant.
The equal-time correlation function of the same field
in a periodic ring with length $L$
is given by
\begin{equation}
  \langle \phi(x) \phi(y) \rangle_L =
        C \left( \frac{\pi}{L} \right)^{2h}
            \frac{1}{(\sin{\pi |x-y|/L})^{2h}} .
\end{equation}
Thus the minimum of the correlation function at $|x-y| = L/2$
is given by $C (\pi/L)^{2h}$.
In Fig.~\ref{fig:fits} we plot the value of
correlation function at the middle point
measured in our Monte Carlo calculation.
(We averaged the central $7$ --$11$ sites for $ L \geq 80$.)
We compare it with the prediction from the conformal invariance
with parameters $C=0.12$ and $2h=1.3$,
which are determined from the fitting~(\ref{eq:powerlaw}).
Although the data fit well with the prediction for $L \leq 100$,
we see the data deviate upward for $L \geq 120$.
This suggests that the string correlation converges to a finite value
in the thermodynamic limit, which means the
hidden $Z_2 \times Z_2$ symmetry
is broken (at least partially) at this point.

We find a similar result for $D=0.9$
although the string correlation
function is smaller than that for $D=1.2$.
This result also supports that
the phase diagram for $S=2$ is different from that for $S=1$
where there is a single critical point $D=D_c$.
Thus we conclude that there exists the
conjectured intermediate-$D$ phase with a finite region
between the Haldane phase and the large-$D$ phase.

To summarize, we performed a Monte-Carlo calculation on
the $S=2$ quantum antiferromagnetic spin chain with an anisotropy
as in eq.~(\ref{eq:Dham}).
We found that, for $D=1.2$,
the den Nijs-Rommelse string order tends to remain finite
(at least decays slower than a simple power-law) in contrast to
the vanishing N\'{e}el order.
Together with the previous numerical and rigorous results that the
string correlation decays exponentially
at the isotropic point $D=0$ and at sufficiently large $D$,
our data shows there is an intermediate phase, which is absent
in the corresponding $S=1$ chain.
This confirms the conjecture based on a VBS-type model
in Ref.~\cite{MOVBS}.

Our result confirms numerically
that, while the presence of the Haldane gap would
be universal for all integer spins,
the phase diagram of  quantum spin chain
is not universal among all integer spins
and would become more complex as the spin quantum
number is increased.
It is also suggested that the VBS-model approach would be qualitatively
useful even for standard Heisenberg-type models with higher spins,
although the quantitative discrepancy increases as spin quantum
number $S$ increases.

The fact that the string correlation behaves like a power-law
suggests that the system resides near a criticality.
Recently it is argued~\cite{Joli} that there is an XY-phase
in the finite-$D$ region; this XY-phase like character
may be related to the power-law-like behavior of the string correlation.
In the present work, however, we
found that the string correlation in $z$-axis deviates
from a simple power-law, implying the spontaneous breakdown of the
hidden $Z_2 \times Z_2$ symmetry.
This rather suggests that the system is not in a truly gapless XY-phase
in the thermodynamic limit.
In any case,
the intermediate phase studied in this letter
can be distinguished from the XY phase in the ferromagnetic
region (or equivalently with a negative $S^z_i S^z_{i+1}$ coupling)
by the dominance of the string order.

M.O. thanks Hal Tasaki and Naoto Nagaosa for stimulating discussions.
M.O. and M.Y. are grateful to Mahito Kohmoto and Yasuhiro Hatsugai
for useful discussions on various related topics.
The calculation is done on HP 9000 715 workstation, HITAC S3800
supercomputer at the Computer Center of University of Tokyo and
VPP500 supercomputer at the Institute for Solid State Physics,
University of Tokyo.
The present work is partly supported by Grant-in-aid for Scientific
Research on Priority Areas ``Novel Electronic States in
Molecular Conductors'', from Japanese Ministry of Education,
Science and Culture.

\end{document}